\documentclass[conference]{IEEEtran}
\IEEEoverridecommandlockouts
\usepackage{cite}
\usepackage{amsmath,amssymb,amsfonts}
\usepackage{graphicx}
\usepackage{textcomp}
\usepackage{xcolor}
\usepackage{booktabs}
\usepackage{url}
\graphicspath{{figures/}}

\begin{document}

\title{Battery-Buffered Solid-State Transformers for AI Data Centers: Load Smoothing, Shaft Torsion and Fault Ride-Through}

\author{\IEEEauthorblockN{Eihab E. E. Ahmed}
\IEEEauthorblockA{\textit{Electrical and Computer Engineering} \\
\textit{San Diego State University and University of California San Diego}\\
San Diego, CA, USA \\
eelgailyelmuks5457@sdsu.edu}
\and
\IEEEauthorblockN{Saeed Manshadi}
\IEEEauthorblockA{\textit{Electrical and Computer Engineering} \\
\textit{San Diego State University}\\
San Diego, CA, USA \\
smanshadi@sdsu.edu}}

\maketitle

\begin{abstract}
AI training campuses threaten grid stability because their fluctuating load can excite the torsional modes of nearby
generators, and grid codes now require them to ride through faults. We study a battery energy storage system
(BESS) on the 800 V DC bus behind a solid-state transformer (SST) in an electromagnetic transient model. The
BESS buffers sub-second fluctuations; the SST imports the filtered IT demand. A 1-s filter cuts the grid
power's standard deviation by 67\%, its maximum 1-s ramp by 68\%, and the first torsional-mode torque by 95\%.
A BESS at 54\% of the IT peak preserves the standard deviation and ramp rate at the point of interconnection but
raises torsional torque 3.7 to 5 times over an IT-peak BESS. Ride-through needs a BESS near the IT peak,
since the SST blocks its current during the sag. The campus then stays connected at the edge of the ERCOT
ride-through envelope.
\end{abstract}


\section{Introduction}

Training workloads synchronize thousands of accelerators, so a hyperscale AI data center can move hundreds of megawatts between compute and communication phases within a fraction of a second~\cite{Choukse2025PowerStabilization}. These swings can excite inter-area modes~\cite{Ko2026AIWideArea} and accumulate fatigue in turbine-generator shafts~\cite{Hossain2026ThermalFatigue}, system operators now ask for dynamic models of such loads~\cite{esig_large_load_modeling_2026}, and ERCOT mandates their fault ride-through~\cite{ERCOT2026NOGRR282}.

SSTs that convert medium-voltage AC directly to an 800 V DC bus are being adopted for these campuses~\cite{Xu2026SST800V}, and the Open Compute Project (OCP) SST specification sets a DC voltage tolerance and a low-voltage ride-through (LVRT) requirement~\cite{OCP2026SST}. Meeting both without curtailing the IT load during deep sags calls for energy storage, and where that storage sits determines what it can do. Rack-level battery backup units and capacitor banks absorb sub-second steps close to the accelerators~\cite{Choukse2025PowerStabilization}, facility UPS batteries can be coordinated for load flexibility~\cite{Li2026UPSBatteries}, and AC-side grid-forming storage raises the post-fault frequency nadir and damps training-load torque oscillations~\cite{Ross2026GFMESS}. An SST-based campus allows a battery energy storage system (BESS) on the 800 V DC bus behind the SST, where a single unit serves both the IT load and the grid interface.

When the battery sits behind the SST, two concerns arise that AC-side storage studies do not address. First,
the split of power between the grid and the battery becomes a design choice, and its effect on grid-side
variability, ramp, generator shaft torque and battery duty has not been quantified for training workloads. Second, the OCP
zero-current rule transfers the full IT load to the DC-side battery during a deep sag, and the restoration of
grid current after clearing interacts with the DC bus regulation.

This paper addresses both concerns with an OCP-aligned electromagnetic transient (EMT) model built on the OpenIBR library~\cite{OpenIBR800V}, in which the BESS regulates the 800 V bus within explicit power and state-of-charge (SOC) limits while the SST follows a filtered feed-forward of the IT demand. We quantify how the feed-forward time constant trades grid-side variability, ramp and the torsional excitation of a nearby turbine-generator against battery throughput, and show that the shaft sets a stricter battery rating than the grid-side metrics. We then propose an LVRT policy that combines zero-current blocking, a hold of the virtual synchronous machine (VSM) states and a soft current restoration, and an ablation study isolates the contribution of each function.

\section{System and Model}

\begin{figure}[!t]
\centering
\includegraphics[width=\columnwidth]{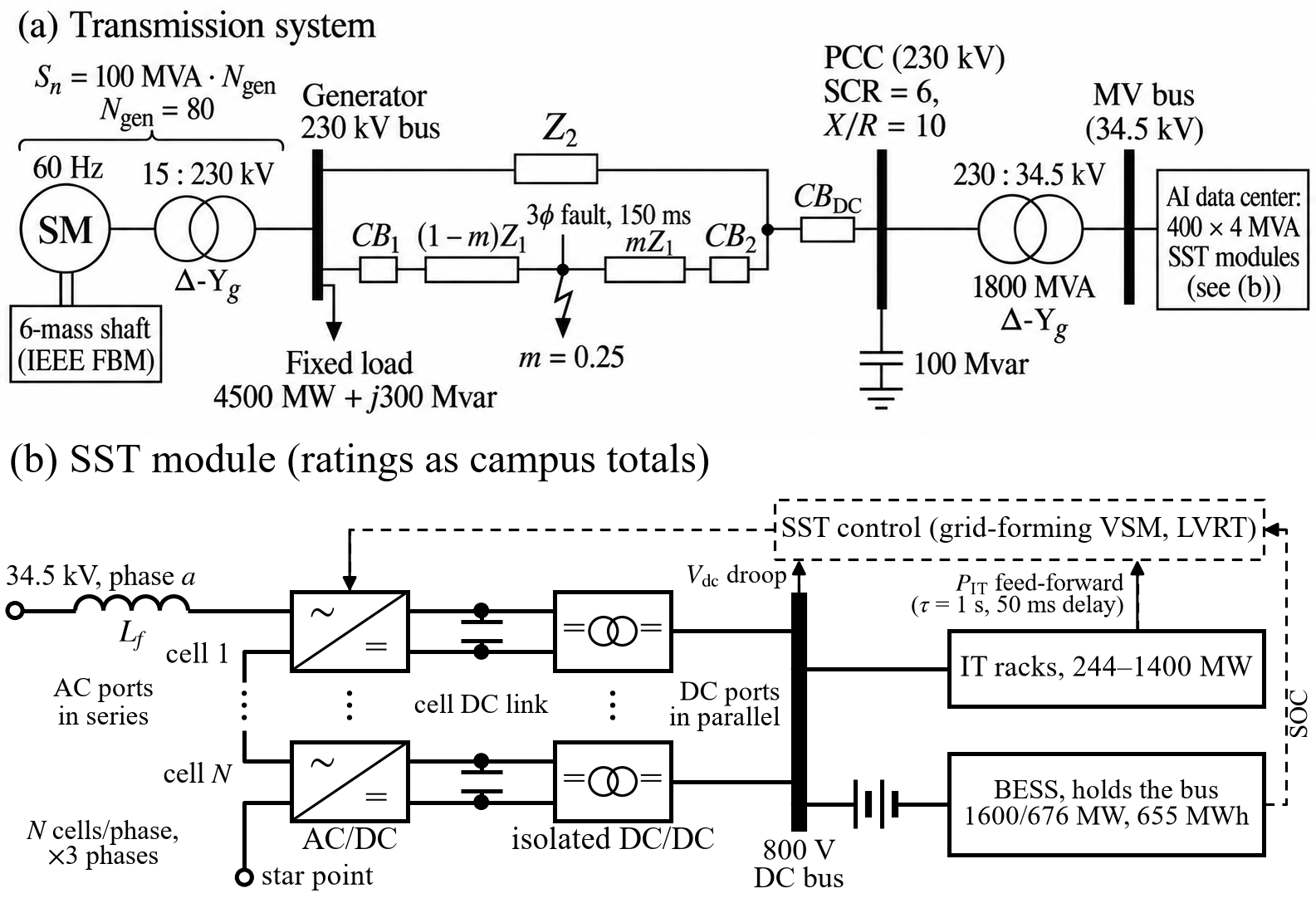}
\caption{Study system: (a) transmission system and campus connection; (b) one SST module, with ratings given as
campus totals. Each of the 400 SST modules feeds its own 800 V DC bus with an IT load and a battery; the modules
are aggregated by current scaling.}
\label{fig:system}
\end{figure}

\subsection{Grid and Load}
Fig.~\ref{fig:system} shows the study system, implemented in PSCAD/EMTDC. To allow direct comparison with
AC-side storage, the transmission system and generator follow the test system of~\cite{Ross2026GFMESS}, with a coherent equivalent of 80 identical 100 MVA thermal units (8 GVA) at
15/230 kV, a short-circuit ratio of 6 at the point of common coupling (PCC), an AC1A exciter and a steam governor
with 5\% droop. Each unit carries the six-mass shaft of the IEEE First Benchmark
Model~\cite{IEEE1977FirstBenchmarkSSR}. The generator serves a 4500 MW, 300 Mvar fixed load and, through a
1800 MVA transformer, the data-center campus. 

The campus consists of 400 SST modules rated 4 MVA each (1600 MVA in total). Each module feeds an IT load,
modeled as a constant-power DC load. The load profile is the measured GPU power trace supplied with the
ERCOT/TAMU data-center model~\cite{ERCOT2026DynamicAIDC}, scaled linearly so that the campus demand spans
244--1400 MW, with steps of about 1 GW roughly every 1.5 s. This scaling assumes that all modules
swing in phase and that there is no non-IT load, so it represents a worst case.

The battery stores 655 MWh (1.64 MWh per module). This energy follows from the 1600 MW rating at 2.4C, not
from the smoothing duty, which moves the SOC by less than 0.05\%, and is below the 2-h systems deployed at data
centers~\cite{Tesla2025Megapack}, while minute-scale UPS batteries are not meant for repeated
ramps~\cite{Lyu2026DelayBESS}.

\subsection{SST Model}
The SST follows the averaged OpenIBR model~\cite{OpenIBR800V}. The AC/DC stage is a current-controlled
bridge. Its reference sums a power-request term derived from the battery SOC and the IT
feed-forward, an OpenVSM term that emulates a synchronous machine with an inertia constant of 1 s, and a
regulator that keeps the average of the three 33 kV medium-voltage (MV) DC links at nominal. Isolated DC/DC
stages hold each MV link and supply the 800 V bus. The total current reference is limited to
$I_\mathrm{max}=1.0$~pu; only the reference case without LVRT in Table~\ref{tab:ablation} uses the 1.5-pu
short-term overload of a grid-forming converter that feeds fault current.

Four changes were made relative to the library model to meet the OCP requirements and to allow the BESS to buffer the fast-oscillating AI training load. 1) The MV-link reference was aligned to 33 kV; 2) the IT demand was added as a feed-forward to the
power request; 3) the battery was given explicit power and SOC limits; and 4) the LVRT functions of
Section~\ref{sec:lvrt} were added. Table~\ref{tab:params} lists the main parameters.

\begin{table}[!t]
\centering
\caption{Main Model Parameters (Campus Totals Unless Noted)}
\label{tab:params}
\footnotesize
\begin{tabular}{@{}ll@{}}
\toprule
Parameter & Value \\
\midrule
SST modules, rating & 400 $\times$ 4 MVA \\
Campus transformer & 1800 MVA, 230/34.5 kV \\
IT demand range & 244--1400 MW \\
MV DC links / LV bus & 3 $\times$ 33 kV / 800 V \\
BESS energy & 655 MWh \\
BESS discharge / charge limit & 1600 / 676 MW \\
SOC window, initial SOC & 0.10--0.90, 0.60 \\
Feed-forward $\tau$ & 0.02--5 s (nominal 1 s) \\
\bottomrule
\end{tabular}
\end{table}

\section{Control Design}

The controls are designed to act on three time scales. The BESS regulates the 800 V bus within milliseconds, the SST
follows a filtered IT feed-forward over about a second, and a slow loop restores the SOC over several seconds.
A droop covers battery saturation, and an LVRT gate overrides the SST current during deep sags.

\subsection{Power Split Between Grid and Battery}
In the Laplace domain, the active power that the SST requests from the grid is:
\begin{equation}
P_\mathrm{g}^{*} = \mathcal{L}_{\tau}\big\{e^{-sT_d}P_\mathrm{IT}\big\} + P_\mathrm{soc},
\label{eq:split}
\end{equation}
where $P_\mathrm{IT}$ is the measured IT power of the module, $s$ is the Laplace variable, $T_d=50$~ms the
measurement and communication delay of the IT power, $\mathcal{L}_{T}\{\cdot\}$ a first-order low-pass filter
$1/(1+sT)$ with time constant $T$, $\tau$ the feed-forward time constant, and $P_\mathrm{soc}\ge0$ the SOC
recharge term
\begin{equation}
P_\mathrm{soc} = S_\mathrm{b}\,\mathcal{L}_{T_s}\Big\{\mathrm{sat}_{[0,1]}\big(K_s\,\mathcal{L}_{T_f}\{\mathrm{SOC}_0-\mathrm{SOC}\}\big)\Big\},
\label{eq:soc}
\end{equation}
where SOC is the battery state of charge, $\mathrm{SOC}_0=0.6$ its reference, $K_s=20$ a gain, $T_f=0.2$~s and
$T_s=7.5$~s filter time constants, $S_\mathrm{b}=4$~MVA the module rating, and $\mathrm{sat}_{[a,b]}$ clips its
argument to $[a,b]$. The correction is one-sided; a SOC below $\mathrm{SOC}_0$ raises the grid import by at
most one module rating to recharge the battery, and $T_s$ keeps it slower than the feed-forward. Since the battery regulates the bus voltage, it picks up the rest of the load, $P_\mathrm{b}$, which is positive
during discharge. For an ideal lossless bus with small capacitive energy storage relative to the step size, and
with the grid power
$P_\mathrm{g}\approx P_\mathrm{g}^{*}$,
\begin{equation}
P_\mathrm{b} \approx P_\mathrm{IT} - P_\mathrm{g}
 = \Big(1-\frac{e^{-sT_d}}{1+s\tau}\Big)P_\mathrm{IT} - P_\mathrm{soc}.
\label{eq:pb}
\end{equation}
The battery thus takes the high-pass complement of the workload, with corner frequency $1/(2\pi\tau)$, plus
the mismatch caused by the delay.

\subsection{Battery Regulation of the 800 V Bus}
The battery is modeled as a controlled internal voltage $E$ behind its series resistance. Its control logic is
\begin{multline}
\dot{E} = \max\Big\{\min\Big\{\frac{V^{*}-V_\mathrm{dc}}{T_v},\,
k_I\big(I_\mathrm{d}^{\max}-I_\mathrm{b}\big)\Big\},\\
k_I\big(-I_\mathrm{c}^{\max}-I_\mathrm{b}\big)\Big\},
\label{eq:dcbus}
\end{multline}
where $V_\mathrm{dc}$ is the bus voltage, $V^{*}=800$~V its reference, $T_v=2$~ms the time constant of the
voltage loop, $I_\mathrm{b}$ the battery current (positive when discharging), $k_I=5$~V/(A$\cdot$s) the gain of
the current-limit loops, and $I_\mathrm{d}^{\max}=P_\mathrm{d}^{\max}/V_\mathrm{dc}$
and $I_\mathrm{c}^{\max}=P_\mathrm{c}^{\max}/V_\mathrm{dc}$ are the discharge and charge current limits obtained
from the power limits in Table~\ref{tab:params}. Normally the voltage loop is selected; when a limit is
reached, its loop takes over, and because all loops share the state $E$, none winds up. The discharge limit is
set to zero when the SOC falls to 0.10 and the charge limit when it reaches 0.90.

While a limit binds, the battery no longer holds the bus, and the SST takes over through a droop on its grid
power request,
\begin{equation}
P_\mathrm{g}^{*} \leftarrow P_\mathrm{g}^{*} - K_d\,S_\mathrm{b}\,\mathrm{dz}_{\delta}\big(V_\mathrm{dc}/V^{*}-1\big),
\label{eq:droop}
\end{equation}
where $K_d=10$~pu is the droop gain and $\mathrm{dz}_{\delta}$ a dead zone of $\pm\delta=\pm1$\%, which leaves
normal operation unchanged. Beyond it, the grid absorbs the power that the saturated battery cannot.

\subsection{Low-Voltage Ride-Through}\label{sec:lvrt}
A hysteretic detector sets the block flag $b=1$ when the positive-sequence terminal voltage falls below 0.50 pu
and clears it when the voltage recovers above 0.52 pu. The current reference sent to the inner current loop is
\begin{equation}
i^{*} = r\,\mathrm{sat}_{I_\mathrm{max}}(i_\mathrm{ref}),\quad
r = \begin{cases} 0, & b=1,\\ \min\{1,\,(t-t_\mathrm{c})/T_r\}, & b=0,\end{cases}
\label{eq:lvrt}
\end{equation}
where $i_\mathrm{ref}$ is the current reference from the SST controls, $\mathrm{sat}_{I_\mathrm{max}}$ limits
its magnitude to $I_\mathrm{max}$, $r$ is a restoration factor, and $t_\mathrm{c}$ is the instant at which $b$
last cleared. The AC current is thus zero during a deep sag
(OCP range~2), and after clearing it is restored linearly over $T_r=150$~ms; in normal operation $r=1$. While
$b=1$, the VSM is also disabled, which freezes its states so that they do not wind up against the collapsed
voltage.

\section{Results}

\subsection{Torsional Excitation and Grid-Side Smoothing}\label{sec:smoothing}
\begin{figure}[!t]
\centering
\includegraphics[width=\columnwidth]{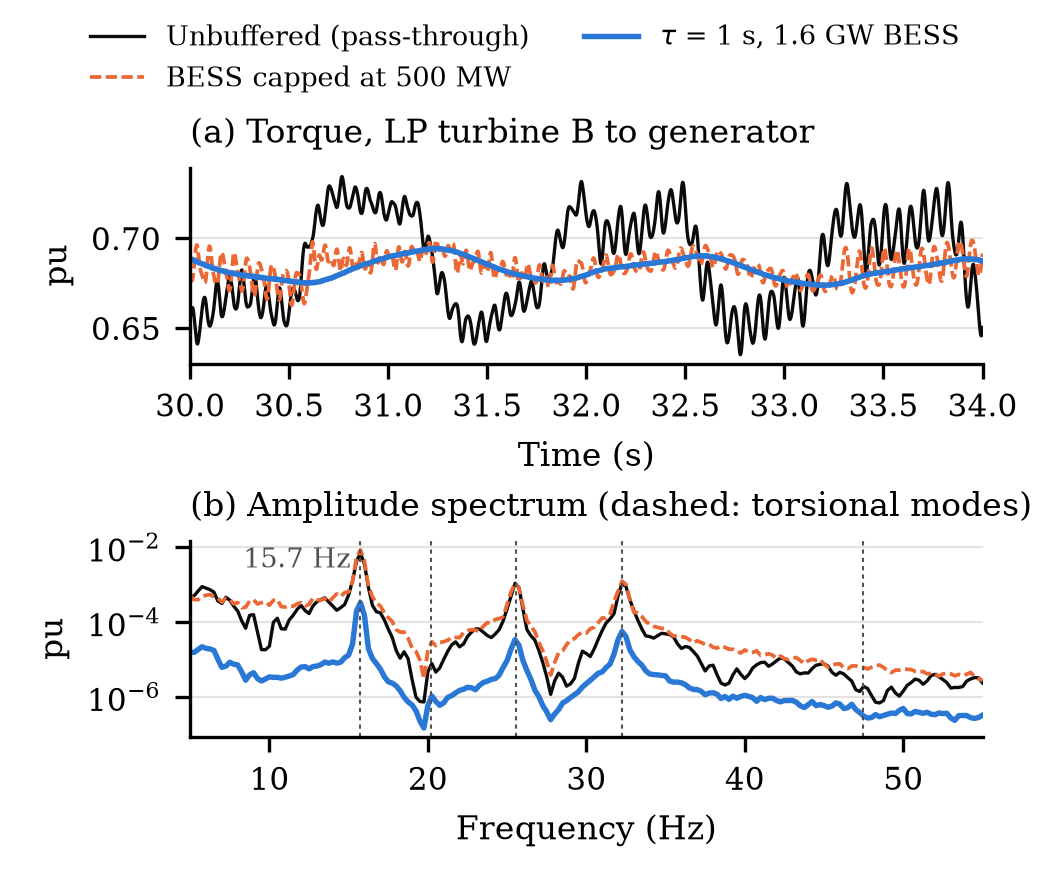}
\caption{Torque in the LPB–generator shaft section: (a) time trace and (b) amplitude spectrum. Dashed lines mark the torsional modes of the benchmark shaft.}
\label{fig:shaft}
\end{figure}

Buffering the workload behind the SST nearly removes the torsional excitation of the nearby generator.
Fig.~\ref{fig:shaft} shows the torque in the shaft section between the second low-pressure (LP) turbine and the
generator. In the unbuffered reference, the SST passes the IT demand straight to the grid ($\tau=1$~ms, no
delay); this load excites all torsional modes of the benchmark shaft, most strongly the 15.7 Hz mode. With the
1.6 GW battery and $\tau=1$~s, the amplitude of this mode falls by 95\% and the RMS torque in the 10--55 Hz band
by 91\%. The other four shaft sections show the same 91--93\% reduction, and the largest oscillation is between the
two LP turbines.

Moreover, even a short filter captures most of this benefit. Five 66-s runs with $\tau\in\{0.02,0.1,0.5,1,5\}$~s and 50 ms feed-forward delay were evaluated over the 51-s workload window from 12 to 63~s, after the model has
initialized and settled at a constant load. Table~\ref{tab:sweep} reports the results of this sensitivity analysis. Relative to the unbuffered row, the $\tau=0.02$~s row thus adds both the 20-ms filter and this delay. At $\tau=0.02$~s the corner frequency is 8~Hz, already below the 15.7 Hz mode, so the
battery removes the sharp step edges that carry the torsional content. Relative to the unbuffered load, the RMS
torque in the torsional band therefore halves while the PCC standard deviation barely changes. At $\tau=0.5$~s the
reduction reaches 88\%, and longer filters add little.

\begin{figure}[!t]
\centering
\includegraphics[width=\columnwidth]{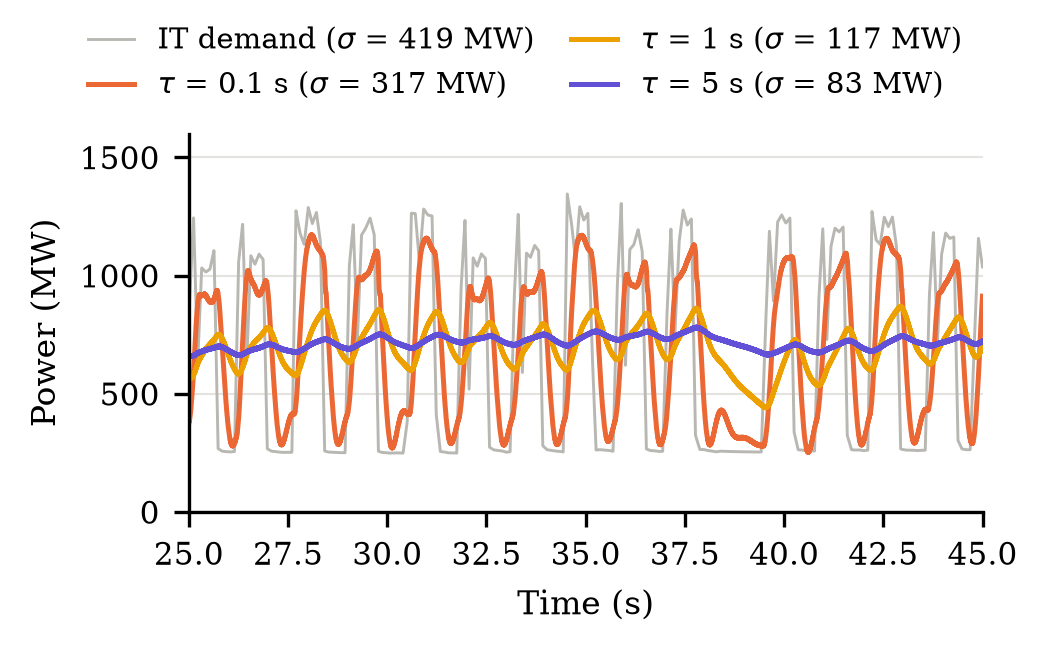}
\caption{PCC active power and IT demand for three feed-forward time constants.}
\label{fig:sweep}
\end{figure}

The grid-side metrics, the standard deviation and the maximum 1-s ramp of the PCC power, by contrast, need
$\tau$ of about 1~s. The workload alternates between peak and idle
power about every 1.5~s, and the PCC power becomes smooth only when the SST import follows more slowly than
these steps. Raising $\tau$ from 0.02 to 5~s lowers the standard deviation of the PCC power from 351 to 83~MW, against
419~MW for the IT demand itself, and the maximum 1-s ramp from 966 to 73~MW/s as shown in Table~\ref{tab:sweep}.
Additionally, Fig.~\ref{fig:sweep} shows the PCC power and standard deviation for three of these time constants.
Most of the reduction occurs between $\tau=0.1$ and 1~s, where the corner frequency falls from 1.6 to 0.16~Hz and
crosses the 0.67 Hz step rate of the workload. Even at $\tau=0.02$~s the PCC standard deviation stays below that
of the IT demand, because the SST loops have finite bandwidth.

The battery pays for the AI load smoothing. In the pass-through case, without smoothing, it already processes 1.55~MWh,
because the SST import lags each load step by its control bandwidth and the battery holds the bus until it catches up. With filtering, its
throughput grows with $\tau$, from 2.51~MWh at $\tau=0.02$~s to 5.51~MWh at $\tau=5$~s over the window, or from
3.3 to 7.1 equivalent full cycles per day if the workload runs continuously; at $\tau=1$~s it is 5.48~MWh, 3.5 times
the pass-through. This throughput counts both charge and discharge, which nearly cancel because the battery returns the energy of each step at the next one. Its SOC therefore swings by less than 0.03\% within a step, drifts by less than 0.01\% over the window, and stays within a 0.05\% peak-to-peak range, so these are very shallow micro-cycles, which cause far less aging than full cycles. Its peak discharge power stays between 810 and 940~MW for every $\tau$, because each workload step is first supplied by
the battery regardless of the filter. At $\tau=0.02$, 0.1 and 0.5~s the 676~MW charge limit binds at the falling steps
for 0.03, 0.98 and 0.25~s, and the droop~\eqref{eq:droop} returns the surplus to the grid, keeping the bus below
817.5~V. At $\tau\ge1$~s no limit binds and the bus stays within $\pm0.1$~V.

\begin{table}[!t]
\centering
\caption{PCC Variation, Shaft Torque and Battery Duty Versus $\tau$ (51-s Window)}
\label{tab:sweep}
\footnotesize
\setlength{\tabcolsep}{1.0pt}
\begin{tabular}{@{}lccccccc@{}}
\toprule
$\tau$ & $\sigma$ & Ramp 1 s & Ramp 10 s & $T_{10\text{--}55}$ & BESS & Thr. & EFC \\
(s) & (MW) & (MW/s) & (MW/s) & ($10^{-4}$) & (MW) & (MWh) & (1/d) \\
\midrule
0.001$^\ast$ & 354 & 974 & 101.6 & 57.7 & $-389$/$417$ & 1.55 & 2.0 \\
0.02 & 351 & 966 & 100.6 & 25.6 & $-703$/$816$ & 2.51 & 3.3 \\
0.1  & 317 & 854 & 88.5 & 19.7 & $-767$/$924$ & 3.25 & 4.2 \\
0.5  & 168 & 502 & 61.0 & 6.7 & $-734$/$940$ & 5.21 & 6.8 \\
\textbf{1} & \textbf{117} & \textbf{315} & \textbf{42.4} & \textbf{5.1} & {\boldmath$-664$/$899$} & \textbf{5.48} & \textbf{7.1} \\
5    & 83  & 73  & 19.8 & 4.7 & $-559$/$810$ & 5.51 & 7.1 \\
\bottomrule
\multicolumn{8}{@{}l}{\scriptsize $^\ast$Unbuffered: no delay. $T_{10\text{--}55}$ in pu. EFC: cycles per day, 655 MWh.}\\
\multicolumn{8}{@{}l}{\scriptsize Peaks include limit-loop overshoot beyond the 676 MW charge limit.}
\end{tabular}
\end{table}

\subsection{Battery Size and Feed-Forward Delay}
Generator shaft oscillations, 800 V bus voltage, and the campus ramp-rate at the PCC set the battery rating for smoothing. Table~\ref{tab:constrained} varies the
battery and the feed-forward delay at $\tau=1$~s; only the discharge limit is reduced, except at 500~MW, where the
charge limit is also 500~MW. A delay of 0--200~ms has a minor effect on the BESS dispatch, and capping the
battery at 1000~MW changes nothing because the discharge power never exceeds 899~MW, or 64\% of the IT peak,
which is thus the rating the shaft requires. At 750~MW (54\%) the discharge limit binds for only 0.4~s, at
the rising steps, and the grid-side metrics barely move, yet the few edges that reach the grid raise the
torsional-band torque 3.7--5 times above the baseline across the shaft sections. The battery briefly reaches
851~MW because the current-limit loop in~\eqref{eq:dcbus}, with finite gain $k_I$, needs up to 54~ms to pull the
current back after a fast step. At 500~MW the discharge and charge limits bind for 12~s of the window in total,
so the edges of each step reach the grid, the 1-s ramp rises from 315 to 384~MW/s, and the torque exceeds the
unbuffered level (Fig.~\ref{fig:shaft}). The droop holds the bus within 755--824~V; without it the battery is
driven beyond its rating, the bus swings between 691 and 943~V, and the delivered IT power falls to 98.4\% of
demand. Its torque is lower (61.1 against 71.9) because the overdriven battery absorbs more of each step edge, which the droop otherwise passes to the grid.

\begin{table}[!t]
\centering
\caption{Battery Size and Feed-Forward Delay ($\tau=1$~s Unless Noted)}
\label{tab:constrained}
\scriptsize
\setlength{\tabcolsep}{1.7pt}
\begin{tabular}{@{}lcccccc@{}}
\toprule
Case & $\sigma$ & Ramp 1 s & BESS & Limit & $V_\mathrm{dc}$ & $T_{10\text{--}55}$ \\
 & (MW) & (MW/s) & (MW) & (s) & (V) & ($10^{-4}$ pu) \\
\midrule
Unbuffered$^\dagger$ & 354 & 974 & $-389$/$417$ & 0 & 800 & 57.7 \\
\textbf{Baseline (1600 MW)} & \textbf{117} & \textbf{315} & {\boldmath$-664$/$899$} & \textbf{0} & \textbf{800} & \textbf{5.1} \\
Delay 0 ms & 117 & 315 & $-652$/$880$ & 0 & 800 & 5.5 \\
Delay 200 ms & 116 & 315 & $-668$/$896$ & 0 & 800 & 5.2 \\
BESS 1400 MW & 117 & 315 & $-664$/$899$ & 0 & 800 & 5.1 \\
BESS 1000 MW & 117 & 315 & $-664$/$899$ & 0 & 800 & 5.0 \\
BESS 750 MW & 116 & 323 & $-664$/$851$ & 0.4 & 779--800 & 18.8 \\
BESS 500 MW & 103 & 384 & $-595$/$609$ & 12.4 & 755--824 & 71.9 \\
BESS 500 MW, no droop & 111 & 474 & $-642$/$810$ & 19.4 & 691--943 & 61.1 \\
\bottomrule
\multicolumn{7}{@{}l}{\scriptsize $^\dagger$$\tau=1$~ms, no delay. BESS $X$ MW: discharge limit; charge limit 676 MW (500}\\
\multicolumn{7}{@{}l}{\scriptsize MW rows: 500 MW). Peaks include limit-loop overshoot. $T_{10\text{--}55}$: LPB--generator.}
\end{tabular}
\end{table}

\subsection{Fault Ride-Through}
\begin{figure}[!t]
\centering
\includegraphics[width=\columnwidth]{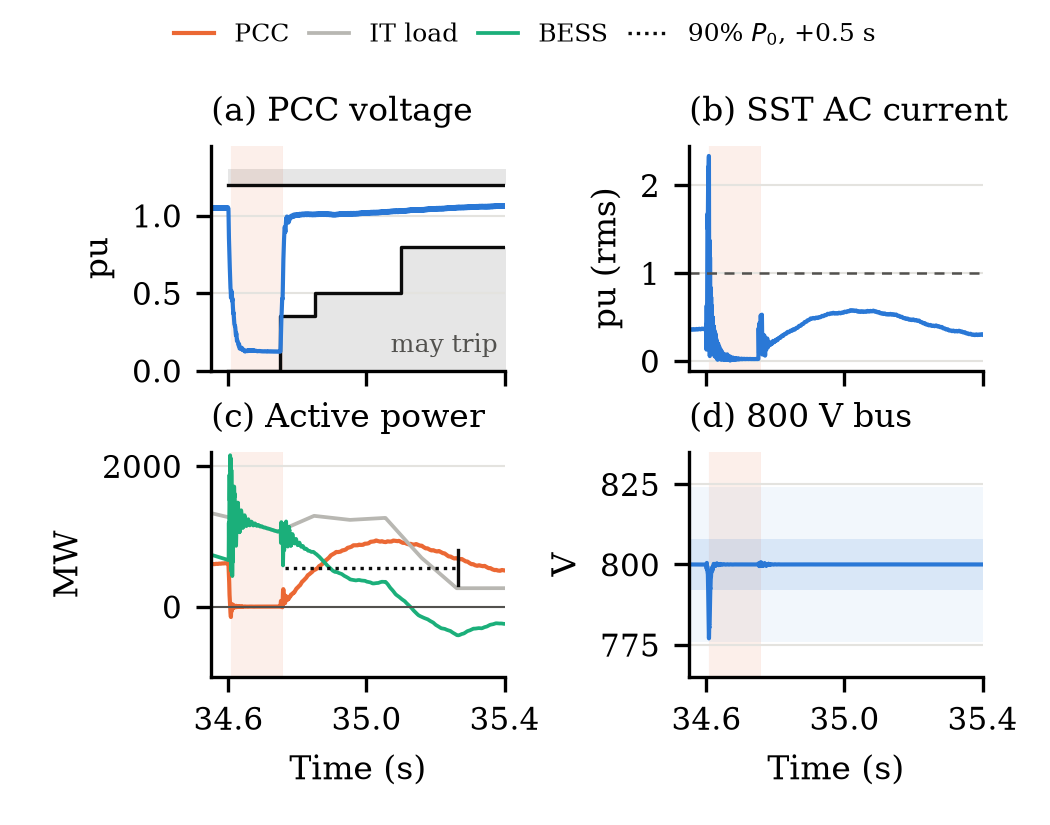}
\caption{Ride-through of a 150-ms three-phase fault at high load with the final LVRT policy. (a) PCC voltage
at the 230 kV service delivery point; gray: region where ERCOT NOGRR~282 allows tripping. (c) Dotted: 90\% of
pre-fault PCC power, required within 0.5~s of recovery. (d) Bands: $\pm1$\% and $\pm3$\%. Shading: AC current
blocked.}
\label{fig:frt}
\end{figure}

With the LVRT functions the campus rides through a deep fault at high load at the edge of the ERCOT envelope,
but only with a battery near the IT peak. A three-phase fault at $t=34.60$~s, cleared after 150~ms, depresses the voltage to about 0.12~pu (OCP
range~2) while the IT demand averages 1164~MW. Fig.~\ref{fig:frt} shows the baseline response. Before
the detector responds after about 7~ms, the SST current reaches 2.3~pu and the battery about 2~GW in every
high-load fault run, regardless of battery size and LVRT functions, because none of them acts before detection. It
exceeds $I_\mathrm{max}$ because the limit acts on the reference, while the collapsing voltage drives the
current faster than the current loop can follow. After detection the AC current is blocked and the battery carries 1157~MW, the full IT
load, within its 1600~MW rating. The 800 V bus dips to 777~V ($-2.9$\%) at the onset and stays within $\pm1$\% afterwards. After clearing,
the current ramps back to a peak of 0.58~pu and the PCC power peaks at 941~MW, against 608~MW before the fault.
The delivered IT power stays within 98.3--101.5\% of demand.

Against the ERCOT envelope~\cite{ERCOT2026NOGRR282}, the PCC voltage stays below
0.35~pu for 138~ms of the required 0.15~s as shown in Fig.~\ref{fig:frt}(a). The campus stays connected and its PCC power regains 90\% of the
pre-fault value 94~ms after recovery, within the 0.5~s allowed. The one exceedance is the onset spike, during which
the PCC current reaches 2.8 times its normal maximum for 12~ms, above the 150\% limit. The averaged OpenIBR model
omits the sub-millisecond hardware protection, such as desaturation detection and pulse blocking, and thus
exposes the worst-case transient during the 7-ms detection delay; in hardware, this protection acts within
about a millisecond.

The battery size that suffices for smoothing does not suffice here. With 1000~MW the battery cannot carry the
blocked load. The bus falls to 698~V ($-12.8$\%), stays outside $\pm3$\% for 208~ms, and the delivered IT power
drops to 96.2\% of demand. With 750 and 500~MW the bus falls to 676 and 627~V, and the droop drives a
post-fault SST current of 2.6 and 3.0~pu. As at the onset, this exceeds $I_\mathrm{max}$ because the limiter
acts on the reference, while the collapsed bus pulls the measured current above it. A fault at low load ($t=30$~s, 250~MW) is ridden through with 500~MW.

An ablation study at the high-load fault isolates the contribution of each LVRT function.
Table~\ref{tab:ablation} starts from a case without LVRT, in which the SST current stays above its 1.5-pu
overload limit throughout the sag, and adds the functions of~\eqref{eq:lvrt} one at a time. Zero-current
blocking stops this current, but the VSM keeps running against the
collapsed voltage, winds up, and releases a post-fault surge of 2.3 times the pre-fault PCC power. Freezing the
VSM states during blocking lowers the surge to 1.7 times, and ramping the current back over 150~ms lowers it
further to 1.5 times.
The LVRT functions thus shape the grid side, not the DC bus. In every case the bus dips by at most 3.9\% at the
fault onset and then stays within $\pm1$\%, because the battery and the droop absorb the surge.

\begin{table}[!t]
\centering
\caption{LVRT Ablation at the High-Load Fault (Each Row Adds One Function)}
\label{tab:ablation}
\footnotesize
\setlength{\tabcolsep}{3pt}
\begin{tabular}{@{}lccccc@{}}
\toprule
Case & $I_\mathrm{sag}$ & $I_\mathrm{post}$ & $P_\mathrm{PCC}^\mathrm{pk}$ & $V_\mathrm{dc}^\mathrm{min}$ & ${>}\pm1$\% \\
 & (pu) & (pu) & (MW) & (V) & (ms) \\
\midrule
No LVRT$^\dagger$ & 1.56 & 1.76 & 1673 & 769 & 8.0 \\
+ Block & 0.03 & 1.03 & 1369 & 777 & 5.8 \\
+ VSM hold & 0.03 & 0.66 & 1063 & 777 & 5.8 \\
\textbf{+ Soft restoration} & \textbf{0.03} & \textbf{0.58} & \textbf{941} & \textbf{777} & \textbf{5.8} \\
\bottomrule
\multicolumn{6}{@{}l}{\scriptsize $^\dagger$$I_\mathrm{max}=1.5$~pu (overload), else 1.0~pu. Pre-fault PCC 608 MW.}\\
\multicolumn{6}{@{}l}{\scriptsize $I_\mathrm{sag}$: mean after the onset.}
\end{tabular}
\end{table}

\section{Conclusion}
This paper investigates a BESS on the 800 V DC bus behind the SST of an AI training campus. An EMT model with a six-mass turbine-generator shaft was used to evaluate the feed-forward time constant, the battery rating and the LVRT policy against the OCP and ERCOT requirements. The results illustrate that the feed-forward time constant sets what the grid sees. With a 1-s filter, the first torsional-mode torque decreased by 95\% and the maximum 1-s ramp by 68\%, at 3.5 times the battery throughput of the unbuffered pass-through. In addition, each requirement calls for a larger battery. A 750 MW battery preserves the grid-side metrics, the shaft requires the 899 MW peak discharge of the 1-s filter (64\% of the IT peak), and ride-through requires the full IT load, because the SST blocks its current during the sag, and therefore sets the rating. It was also found that blocking the SST current by itself winds up the VSM and causes a post-fault surge of 2.3 times the pre-fault power. Freezing the VSM states and ramping the current back reduce this surge to 1.5 times. The study is designed as a worst-case synchronous stress test, with all 400 SST modules swinging in phase and no non-IT load, so the reported shaft torques and DC-bus deviations bound what a host grid and campus would experience. The bus regulation, the smoothing and the SOC recovery act on separate time scales, so the framework extends to longer simulations that count shaft fatigue cycles with the rainflow method and enforce minute-scale utility ramp limits.

\end{document}